\documentclass[twocolumn,showpacs,preprintnumbers,amsmath,amssymb,apl]{revtex4-1}

\usepackage{graphicx}
\usepackage{dcolumn}
\usepackage{bm}
\usepackage{amsmath}
\input{epsf}

\begin{document}

\newcommand{\note}[1]{\textbf{#1}}

\title{Josephson vortex coupled to a flux qubit}
\author{\firstname{Kirill G.} \surname{Fedorov$^{1,2}$}}
\email{kirill.fedorov@kit.edu}
\author{\firstname{Anastasia V.} \surname{Shcherbakova$^{1}$}}
\author{\firstname{Roland} \surname{Sch\"{a}fer$^{3}$}}
\author{\firstname{Alexey V.} \surname{Ustinov$^{1,2,4}$}}
\affiliation{ $^{1}$Physikalisches Institut and DFG-Center for Functional Nanostructures (CFN), Karlsruhe Institute of Technology, D-76131 Karlsruhe, Germany \\
$^{2}$ National University of Science and Technology MISIS,
Leninsky prosp. 4, Moscow 119049, Russia \\
$^{3}$Institut f\"{u}r Festk\"{o}rperphysik, Karlsruhe Institute of
Technology, D-76131 Eggenstein-Leopoldshafen, Germany\\
$^{4}$ Russian Quantum Center, 100 Novaya St., Skolkovo, Moscow region, 143025, Russia  }

\date{\today}

\begin{abstract}
Experiments towards realizing a readout of superconducting qubits by using ballistic Josephson vortices are reported. We measured the microwave radiation induced by a fluxon moving in an annular Josephson junction. By coupling a flux qubit as a current dipole to the annular junction, we detect periodic variations of the fluxon's oscillation frequency versus magnetic flux through the qubit. We found that the scattering of a fluxon on a current dipole can lead to the acceleration of a fluxon regardless of a dipole polarity. We use the perturbation theory and numerical simulations of the perturbed sine-Gordon equation to analyze our results.
\end{abstract}



\maketitle

We are experimentally investigating a new type of detector that has been theoretically proposed for very fast and weakly perturbing readout of superconducting qubits \cite{FluxQubit1,FluxQubit2}. The detection principle is based on measuring a delay time of a ballistic Josephson vortex (fluxon), moving in a Josephson transmission line (JTL) \cite{FR-Averin,FR-Herr}. The vortex propagation delay depends on the state of the qubit
magnetically coupled to the line. Expected advantages of this method are high time resolution and weak perturbation of the qubit. Potentially, the method might be suitable for implementing quantum feedback schemes \cite{QFeedback,Siddiqi-2012} for flux qubits using on-chip fluxon readout.
This approach also offers an opportunity to incorporate the existing single flux quantum logic (SFQ) solutions in the readout, bringing the dream about scalable quantum computer closer to reality \cite{QQ1,RSFQ}.

A Josephson vortex in underdamped JTL has properties of a relativistic particle carrying a magnetic flux quantum $\Phi_0=h/2e$ \cite{Fulton}. The size of a vortex can vary from few to several hundreds of microns, depending on the critical current density $j_c$ and its velocity $u$ inside the junction. By applying a bias current, the vortex can be accelerated up to the Swihart velocity $c_S$, which is the speed of light in JTL. The dynamical properties of a fluxon resemble a classical particle with a well-defined mass and velocity. Nevertheless, at sufficiently low temperatures, quantum properties of fluxons such as tunneling and energy level quantization have been already observed \cite{QFluxon}.

\begin{figure}
        \begin{center}
        \includegraphics[width=\linewidth,angle=0,clip]{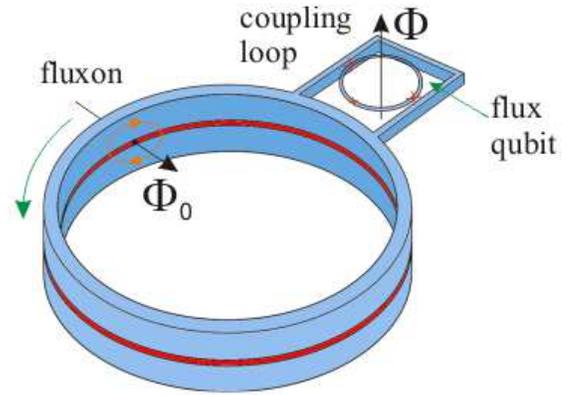}
        \end{center}
    \caption{An annular Josephson junction with a trapped fluxon coupled to a flux qubit.}
    \label{AJJ+Qubit}
\end{figure}

We would like to employ fluxons for developing a fast and sensitive magnetic field detector for measurements of superconducting qubits. In this Letter, we report direct measurements of electromagnetic radiation from a fluxon moving in an annular Josephson junction (AJJ). The radiation is detected by using a microstrip antenna capacitively coupled to the AJJ. Furthermore, we place a flux qubit close to the long junction and couple them magnetically with a superconducting loop (see Fig.~\ref{AJJ+Qubit}). This coupling scheme makes the fluxon interact with a current dipole \cite{MalUst2004,Gold2004} formed by the electrodes of the loop coupled to the qubit. The time delay of the fluxon can be detected as a frequency shift of the electromagnetic radiation emitted from the junction. This shift provides information about the state of the flux qubit.
\begin{figure}
        \begin{center}
        \includegraphics[width=\linewidth,angle=0,clip]{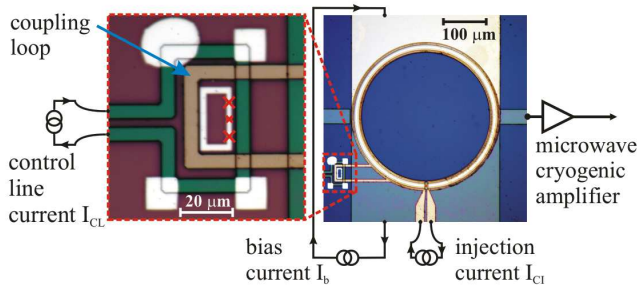}
        \end{center}
    \caption{Optical photograph of the chip with the annular Josephson junction on the right part and experimental
    set-up schematics. Left part shows the zoom into the area with the flux qubit with a coupling loop (yellow loop) and control line (green loop).
    Red crosses indicate the positions of three Josephson junctions in the flux qubit loop.}
    \label{Set-up}
\end{figure}

The most significant advantage of using the closed topology of JTL (see Fig.~\ref{Set-up}) is the quantization of magnetic flux in AJJ, which allows, by putting a pair of current injectors in the biasing electrode, to create a fluxon on demand by applying a current through the injectors \cite{FluxonInsert}.

The circuit was fabricated using photolithography and standard Nb/AlO$_x$/Nb trilayer process with the critical current density $j_c \simeq 1$ kA/cm$^2$ \cite{Hypres}. The estimated Josephson penetration depth is $\lambda_J \simeq 12\, \mu$m, the Josephson plasma frequency $\omega_p/2\pi \simeq 124$ GHz, and the estimated damping parameter $\alpha \simeq 0.02$. The circumference of the junction $L = 1130~\mu$m determines the
frequency of the radiation corresponding to a single fluxon moving with the Swihart velocity $c_S$ to be at about $15$ GHz. The width of the AJJ was $W = 2 \mu$m and its fluxon free critical current $I_c = 23$ mA. The flux qubit was made using the standard aluminum shadow evaporation process \cite{FluxQubit1,ShadowEv} and deposited after the niobium structures were fabricated. Estimated parameters for the Josephson junctions in the flux qubit loop were the following: critical current $I_c = 380$ nA, alpha factor $\alpha_{q} = 0.54$, ratio of Josephson and charging energies $E_J / E_C = 830$.

\begin{figure}
        \begin{center}
        \includegraphics[width=1\linewidth,angle=0,clip]{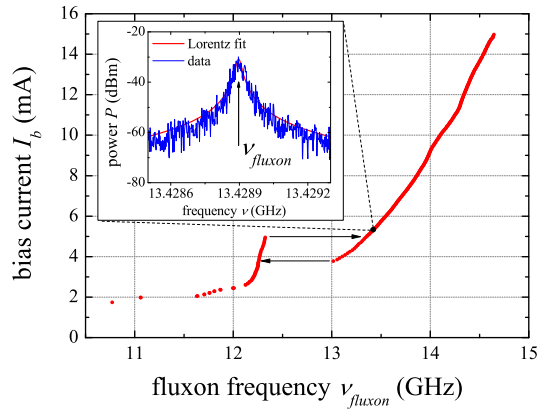}
        \end{center}
    \caption{The zero-field step measured in the frequency domain for the ambient
    temperature $T = 4.2$ K and injection current $I_{CI} = 3.973$ mA. The inset shows the sample spectrum for the fixed bias current $I_b = 5.3$ mA with the respective Lorentz fit.}
    \label{ZFS}
\end{figure}

The fluxon radiation was detected first at $T = 4.2$ K temperature using of a cryogenic wide band (4-20 GHz) microwave amplifier with the noise temperature of about $7$ K followed by a room temperature amplifier with the total gain of 50 dB. The radiation spectrum was studied using a Rohde$\&$Schwarz FSUP26 spectrum analyzer. An example of the measured spectrum is shown in the inset of Fig.~\ref{ZFS}.
Using the possibility to directly detect radiation of the fluxon resonant oscillations, we have performed systematic measurements of the dependence of the fluxon velocity versus bias current - the current-voltage characteristics \cite{Barone} - measured in the frequency domain (see Fig.~\ref{ZFS}). This approach provides an easy access to study the fine structure of the current-voltage curve as the precision of frequency measurements is by several orders of magnitude greater than the resolution of direct dc voltage measurement.

To couple a flux qubit to the fluxon inside an annular Josephson junction, it is necessary to engineer an interaction between two orthogonal magnetic dipoles. To facilitate this interaction, we have added a superconducting coupling loop embracing a flux qubit, as shown in Fig.~\ref{AJJ+Qubit}. The current induced in the coupling loop attached to the AJJ is proportional to the persistent current in the flux qubit. Thus, the persistent current
in the qubit manifests itself in the AJJ as a current dipole with an amplitude $\mu$ on top of the homogeneous background of bias current. When fluxon scatters on a positive current dipole - it first gets accelerated and then decelerated by the dipole poles. In the ideal case of absence of damping and bias current, the sign of frequency change $\delta\nu$ is determined only by polarity of the dipole. In the presence of finite damping and homogeneous bias current, situation completely changes - as the total propagation time becomes dependent on the complex interplay between bias current, current dipole strength and damping.

A theoretical description of interaction between Josephson vortex and current dipole in the AJJ can be done by the perturbed sine-Gordon equation (PSGE) \cite{MalUst2004,Gold2004,McLScott}
\begin{equation}
\frac{\partial^2\varphi}{\partial
t^2}+\alpha\frac{\partial\varphi} {\partial
t}-\frac{\partial^2\varphi}{\partial x^2}=\gamma-\sin
(\varphi)+\mu(\delta(x-d/2)-\delta(x+d/2)) \label{PSGE}
\end{equation}
with the periodic boundary conditions
\begin{equation}
\varphi(-l/2,t) = \varphi(l/2,t) + 2\pi n,
\frac{\partial\varphi(-l/2,t)}{\partial x}=
\frac{\partial\varphi(l/2,t)}{\partial x} \label{Bndr}
\end{equation}
where $n$ is the number of trapped fluxons, $\gamma = I_b / I_c$ is the normalized bias current, $\alpha = \omega_p / \omega_c$ is the damping parameter, $l = L / \lambda_J$ is the normalized junction circumference, $\mu = I_{\mu} / (j_c \lambda_J W)$ is the amplitude of the current dipole and $d = D / \lambda_J$ is the normalized distance between the dipole poles. Direct analytic solution of (\ref{PSGE}) is not an easy task. Therefore, we analyze Eq.~(\ref{PSGE}) using the perturbation approach developed in \cite{McLScott}. In the limit of small perturbations $\gamma \ll 1, \alpha \ll 1$ and $\mu \ll 1$, motion of a single fluxon in the AJJ can be described by a system of ordinary differential equations for the fluxon's velocity $u(t)$ and its spatial coordinate $X(t)$:
\begin{align}
\frac{d u}{d t} = - \frac{\pi\gamma}{4} (1-u^2)^{3/2} - \alpha u
(1-u^2) - \frac{\mu}{4} (1-u^2) \times \nonumber\\ \times \left[ \rm{sech}
\frac{d/2-X}{\sqrt{1-u^2}} - \rm{sech} \frac{-d/2-X}{\sqrt{1-u^2}} \right], \label{FKE1} \\
\frac{d X}{d t} = u - \frac{\mu u}{4} \bigg[ (d/2-X) \rm{sech} \frac{d/2-X}{\sqrt{1-u^2}} + \nonumber\\
+ (d/2+X) \rm{sech} \frac{-d/2-X}{\sqrt{1-u^2}} \bigg]. \label{FKE2}
\end{align}
\begin{figure}
        \begin{center}
        \includegraphics[width=\linewidth,angle=0,clip]{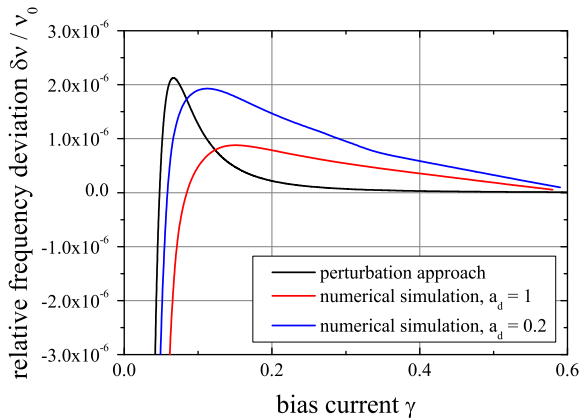}
        \end{center}
    \caption{Relative frequency deviation from equilibrium $\delta \nu / \nu_0$ of the fluxon oscillation frequency versus bias current. Black line shows the result of perturbation approach, while the red line depicts results of direct numerical simulations of the PSGE equation (\ref{PSGEm}) with $a_d = 1$. The blue curve corresponds to the case with $a_d = 0.2$.}
    \label{FD}
\end{figure}

By numerically solving (\ref{FKE1})-(\ref{FKE2}) with the additional condition $u(-l/2) = u(l/2)$ one can calculate an equilibrium trajectory in phase space for fluxon oscillations in the AJJ with the current dipole and estimate a deviation of fluxon oscillation frequency from the unperturbed case $\delta \nu = \nu_{\mu} - \nu_0$, where $\nu_0$ is the oscillation frequency for $\mu = 0$. Black line in Fig.~\ref{FD} shows the dependence of relative deviation $\delta\nu / \nu_0$ versus bias current $\gamma$ calculated from the perturbation theory for the following set of system parameters: $l = 20$, $\alpha = 0.02$, $\mu = 0.05$, $d = 2$. The deviation $\delta\nu$ is large and negative for small bias currents $\gamma \ll 0.1$, what means that the fluxon is being slowed down by the current dipole and eventually can be pinned at the dipole if the bias current is too small. Surprisingly, for larger currents $\gamma > 0.05$ the sign of $\delta\nu$ becomes positive meaning that the current dipole accelerates the fluxon. To understand this phenomenon, we need to look at the Eq.~(\ref{FKE1}) and notice that the effective damping term $\alpha_e = \alpha u (1-u^2)$ has a non-monotonic behavior. When increasing the fluxon velocity $u$, the effective damping is increasing for $u \le 1 / \sqrt{3}$ and then starts decreasing. This means that deceleration (acceleration) is favorable for low (high) bias currents.

To verify the results of the perturbation theory, we performed direct numerical simulation of Eq.~(\ref{PSGE}) with delta functions replaced by the hyperbolic secants in order to smoothen current distribution:
\begin{align}
\frac{\partial^2\varphi}{\partial t^2}+\alpha\frac{\partial\varphi} {\partial
t}-\frac{\partial^2\varphi}{\partial x^2} = \gamma-\sin (\varphi) + \nonumber \\
+\frac{\mu}{\pi a_d}\left[{\rm sech} \left(\frac{x-d/2}{a_d}\right)-{\rm sech} \left(\frac{x+d/2}{a_d}\right) \right]. \label{PSGEm}
\end{align}
The parameter $a_d$ characterizes the width of current distribution and is $a_d \sim 1$ in the experiment. The prefactor $B = 1 / (\pi a_d)$ is chosen to keep the normalization constraint $B \int^{+\infty}_{-\infty} {\rm sech} ((x-d/2)/a_d) dx = 1$. The red curve in Fig.~\ref{FD} shows the results of the numerical calculations of Eq.~(\ref{PSGEm}) with boundary conditions (\ref{Bndr}). As it can be seen in Fig.~\ref{FD}, results of the numerical simulations qualitatively coincide with the perturbation theory. We see that this coincidence is improving for smaller $a_d$ as Eq.~(\ref{PSGEm}) takes form of Eq.~(\ref{PSGE}) in the limit of $a_d \rightarrow 0$.

\begin{figure}
        \begin{center}
        \includegraphics[width=\linewidth,angle=0,clip]{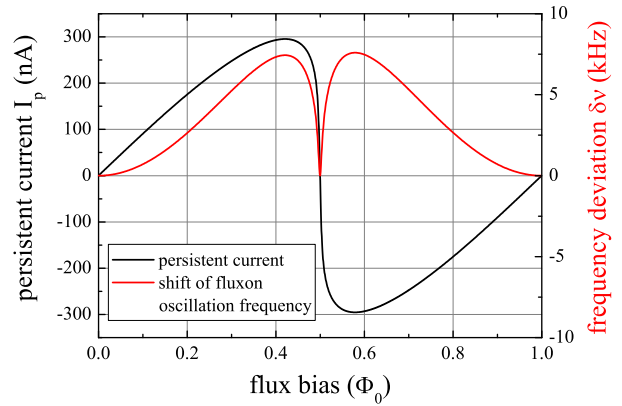}
        \end{center}
    \caption{Persistent current for a ground state of the flux qubit versus magnetic frustration (black line). Red line shows the corresponding fluxon shift calculated using the perturbation theory for.}
    \label{FQ_Icc}
\end{figure}
To experimentally test the qubit readout scheme discussed above the temperature was lowered to $T \simeq 70$ mK, well below the superconducting transition temperature $T_c$ of aluminum forming the qubit. The long junction was biased at a fixed current $I_b$. Then, we varied a current through the control line $I_{CL}$ in order to change the magnetic flux through the flux qubit. Due to the periodic variation of the persistent current in the qubit loop with the control line current $I_{CL}$, the current dipole strength $\mu$ is modulated with an amplitude $\delta\mu$, which depends on the state of the flux qubit. We can write the dipole strength as $\mu = \mu_0 + \delta\mu$, where $\mu_0$ is the initial current offset (e.g., due to trapped magnetic flux). A persistent current $I_p$ of the ground state in the flux qubit can be calculated by numerical simulation of its Hamiltonian \cite{OrlMoo_PRB99}. The result is depicted by the black solid line in Fig.~\ref{FQ_Icc}.

Using the presented perturbation theory (\ref{FKE1})-(\ref{FKE2}) we calculated the response of the fluxon readout to the current dipole controlled by the signal $\mu = k I_p$ (at $\mu_0 = 0$). The proportionality coefficient $k$ is determined by mutual inductance between the qubit and the control loop as well as the critical current density of the AJJ. The fluxon response to the persistent current in the flux qubit loop is depicted in Fig.~\ref{FQ_Icc} by the red curve. It was calculated for a fixed bias current $\gamma = 0.2$, $l = 20$, $\alpha = 0.02$, $\mu(I_p = 300~\rm{nA}) = 0.05$. The response is indicated in kHz assuming $\nu_0 \sim 13$ GHz (as it corresponds in experiment to the bias current $\gamma \sim 0.2$). Noticeably, the response signal $\delta\nu$ is approximately proportional to the amplitude of the persistent current and stays positive despite the change of the sign of $I_p$. The asymmetry of deviation $\delta\nu$ for the positive and negative branches of $I_p$ is less than $3 \%$ . This means that the fluxon scattering is nearly independent of the polarity of the current dipole. It depends dominantly on the bias $\gamma$ and the absolute amplitude of the current dipole $\mu$.

\begin{figure}
        \begin{center}
        \includegraphics[width=\linewidth,angle=0,clip]{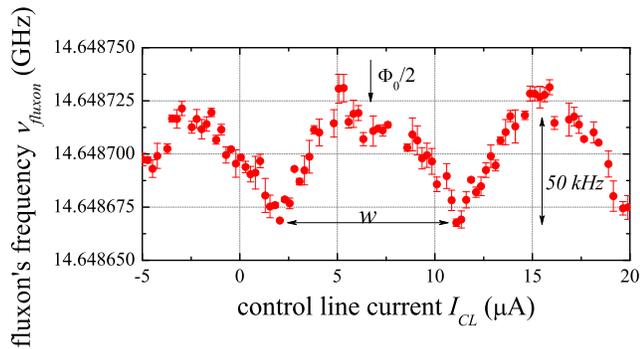}
        \end{center}
    \caption{Modulation of the fluxon's oscillation frequency due to the coupling to the
    flux qubit. Every point consists of $100$ averages. Bias current was set at $\gamma = 0.521$, $w \simeq 9.1$.}
    \label{FM}
\end{figure}
The experimental curve showing the reaction of the fluxon to the magnetic flux through the flux qubit are presented in Fig.~\ref{FM}. The periodic modulation of the fluxon frequency versus magnetic flux through the qubit corresponds to the changing of the persistent currents in the qubit as Fig.~\ref{FQ_Icc} suggests. We did not observe clear narrow peaks at the half flux quantum point, most probably due to excess fluctuations. Emerging dip-like peculiarities can be noted at presumed half flux quantum points which suggest that the dips may be there, covered by noise and insufficient resolution.  Further improvements of experimental setup are required to resolve these peaks. The presented measurement curve has a convex profile which tells that indeed the deviation of frequency $\delta\nu$ is positive, consistently with predictions made above by the perturbation approach and numerical simulations.

In conclusion, we have detected fluxon radiation from the annular Josephson junction at millikelvin temperatures. Measurements of fluxon oscillation frequency as a function of bias current $I_b$ resolve the fine structure of the current-voltage characteristics with much greater precision than direct voltage measurements. Using this technique, we have detected the modulation induced by the persistent current in the flux qubit coupled to the AJJ. We have thus implemented a microwave generator controlled by the flux qubit. We have observed that the scattering of the fluxon on a current dipole can lead to acceleration of the fluxon, regardless of a dipole polarity. The perturbation theory and direct numerical simulations qualitatively well describe this phenomenon. The tested fluxon readout scheme is compatible with Single Flux Quantum (SFQ) superconducting logic and can also be useful for applications where fast, weakly perturbing magnetic signal detection is needed.

The authors would like to acknowledge stimulating discussions with A. Shnirman, A.L. Pankratov, H. Rotzinger and M. Jerger. This work was supported in part by the Ministry of Education and Science of the Russian Federation, the EU project SOLID, the Deutsche Forschungsgemeinschaft (DFG) and the
State of Baden-W{\"u}rttemberg through the DFG Center for Functional Nanostructures (CFN).


\begin{thebibliography}{0}
\bibitem{Hypres} The sample used in this work was fabricated at HYPRES Inc., Elmsford, NY 10523.
\end{thebibliography}
\end{document}